\documentclass[graybox]{svmult}

\usepackage{mathptmx}       
\usepackage{helvet}         
\usepackage{courier}        
\usepackage{type1cm}        
\usepackage{makeidx}         
\usepackage{epsfig}
\usepackage[intlimits]{amsmath}    
\usepackage{amssymb,amsfonts}
\usepackage{graphicx}        
\usepackage{multicol}        
\usepackage[C40,T2A,T1]{CJKutf8}
\usepackage[nocjkjis]{ucs}
\usepackage[bottom]{footmisc}

\renewcommand{\phi}{\varphi}
\renewcommand{\theta}{\vartheta}
\renewcommand{\iota}{\ensuremath{\mathbf{\imath}}}
\renewcommand{\epsilon}{\varepsilon}

\newcommand{\qq}[1]{{\lq}#1{\rq}}

\newcommand{\chn}[1]{\bgroup\begin{CJK*}{UTF8}{}\CJKtilde\CJKfamily{gbsn}{#1}\end{CJK*}\egroup}
\newcommand{\twn}[1]{\bgroup\begin{CJK*}{UTF8}{}\CJKtilde\CJKfamily{bkai}{#1}\end{CJK*}\egroup}
\newcommand{\jap}[1]{\bgroup\fontencoding{C40}\fontfamily{song}\selectfont\SetUnicodeOption{cjkjis}{#1}\egroup}
\newcommand{\ukr}[1]{\bgroup\fontencoding{T2A}\foreignlanguage{ukrainian}{#1}\egroup}
\newcommand{\rus}[1]{\bgroup\fontencoding{T2A}\foreignlanguage{russian}{#1}\egroup}

\newcommand{\LJTS}{LJ{$\cdot$}TS}

\newcommand{\chem}[1]{\ensuremath{\mathsf{#1}}}
\newcommand{\CNT}[1]{\ensuremath{{#1}_\mathrm{CNT}}}
\newcommand{\committorof}[1]{\ensuremath{\committor\left({#1}\right)}}
\newcommand{\contactangle}{\ensuremath{\mathnormal{\theta}}}
\newcommand{\critical}[1]{\ensuremath{{#1}^\star}}

\newcommand{\liq}[1]{\liquid{#1}}
\newcommand{\liquid}[1]{\ensuremath{{#1}'}}

\newcommand{\nuclrateof}[1]{\ensuremath{\nuclrate_{#1}}}

\newcommand{\probinftyof}[1]{\committorof{#1}}

\newcommand{\sat}[1]{{#1}_\mathrm{s}}

\newcommand{\spinodal}[1]{\ensuremath{{#1}^{\#}}}

\newcommand{\surfacetensionof}[1]{\ensuremath{\surfacetension_{#1}}}
\newcommand{\vap}[1]{\vapor{#1}}
\newcommand{\vapor}[1]{\ensuremath{{#1}''}}

\newcommand{\absnum}{\ensuremath{\mathnormal{N}}}

\newcommand{\area}{\ensuremath{\mathnormal{F}}}

\newcommand{\chempot}{\ensuremath{\mathnormal{\mu}}}

\newcommand{\CNTpreexp}{\ensuremath{\consta}}

\newcommand{\collrate}{\ensuremath{\mathsf{T}}}
\newcommand{\committor}{\ensuremath{\mathsf{Q}}}

\newcommand{\consta}{\ensuremath{\mathsf{C}}}
\newcommand{\constb}{\ensuremath{\mathsf{K}}}

\newcommand{\density}{\ensuremath{\mathnormal{\rho}}}

\newcommand{\Deltamu}{\ensuremath{\mathnormal{\Delta\chempot}}}
\newcommand{\Deltamueff}{\ensuremath{\Deltamu_\mathrm{e}}}
\newcommand{\Dgs}{\ensuremath{\Delta\surfacetension_\mathrm{s}}}

\newcommand{\differential}{\ensuremath{\mathnormal{d}}}

\newcommand{\distance}{\ensuremath{\mathnormal{r}}}

\newcommand{\distancecut}{\ensuremath{\distance_\mathrm{c}}}

\newcommand{\FWenergy}{\ensuremath{\mathnormal{\zeta}}}
\newcommand{\FWrect}{\ensuremath{\FWenergy_\mathnormal{0}}}

\newcommand{\helmholtz}{\ensuremath{\mathnormal{A}}}

\newcommand{\inta}{\ensuremath{\mathnormal{\iota}}}
\newcommand{\intb}{\ensuremath{\mathnormal{j}}}

\newcommand{\LJenergy}{\ensuremath{\mathnormal{\varepsilon}}}
\newcommand{\LJenergyFW}{\ensuremath{\LJenergy_\mathrm{fw}}}
\newcommand{\LJlength}{\ensuremath{\mathnormal{\sigma}}}
\newcommand{\LJlengthFW}{\ensuremath{\LJlength_\mathrm{fw}}}

\newcommand{\lsone}{\textit{ls$1$}}

\newcommand{\mardyn}{\lsone{} \textit{mardyn}}
\newcommand{\molarfraction}[1]{\ensuremath{\mathnormal{y}_\mathnormal{#1}}}

\newcommand{\monomers}{\ensuremath{\absnum'}}

\newcommand{\mult}{\ensuremath{\cdot}}

\newcommand{\NcritCNT}{\ensuremath{\nuclsize^\star}}
\newcommand{\nonaccommodation}{\ensuremath{\mathsf{N}}}

\newcommand{\nuclrate}{\ensuremath{\mathnormal{J}}}
\newcommand{\nuclsize}{\inta}

\newcommand{\planartension}{\surfacetensionof{\infty}}

\newcommand{\potential}{\ensuremath{\mathnormal{u}}}

\newcommand{\pressure}{\ensuremath{\mathnormal{p}}}

\newcommand{\radius}{\ensuremath{\mathnormal{R}}}
\newcommand{\radequimolar}{\ensuremath{\radius_\density}}
\newcommand{\radlaplace}{\ensuremath{\radius_\mathsf{L}}}
\newcommand{\radmin}{\ensuremath{\radius_\mathrm{min}}}

\newcommand{\radstillinger}{\ensuremath{\radius_\mathnormal{\ell}}}

\newcommand{\redDgs}{\ensuremath{\constb_\surfacetension}}

\newcommand{\surfacetension}{\ensuremath{\mathnormal{\gamma}}}
\newcommand{\supersat}{\mathnormal{S}}
\newcommand{\supersatmu}{\ensuremath{\supersat_\chempot}}

\newcommand{\supersatrho}{\ensuremath{\supersat_\density}}
\newcommand{\supfraction}{\ensuremath{\mathnormal{\molarfraction{0}}}}
\newcommand{\temperature}{\ensuremath{\mathnormal{T}}}

\newcommand{\threshold}{\ensuremath{\mathnormal{\ell}}}

\newcommand{\tolmanlength}{\ensuremath{\mathnormal{\delta}}}
\newcommand{\tolmaninf}{\ensuremath{\tolmanlength_\infty}}

\newcommand{\virialcoefficient}{\ensuremath{\mathnormal{b}}}

\newcommand{\volume}{\ensuremath{\mathnormal{V}}}
\newcommand{\Weffect}{\ensuremath{\mathsf{W}}}

\newcommand{\zeldovich}{\ensuremath{\mathnormal{z}}}

\newcommand{\demon}{d\ae{}mon}

\newcommand{\etal}{\textit{et al.}}

\renewcommand{\textsc}[1]{{#1}}

\newcommand{\figwidth}{7.5cm}

\begin{document}

\title*{Static and dynamic properties of curved vapour-liquid interfaces by massively parallel molecular dynamics simulation}
\titlerunning{Curved vapour-liquid interfaces by massively parallel MD simulation}
\author{M.\ T.\ Horsch \and S.\ K.\ Miroshnichenko \and J.\ Vrabec\thanks{Corresponding author: Prof.\ Dr.-Ing.\ habil.\ Jadran Vrabec, Universit\"at Paderborn. E-mail: jadran.vrabec@uni-paderborn.de; phone: +49 5251 60 2421; fax: +49 5251 60 3522.} \and C.\ W.\ Glass \and C.\ Niethammer \and M.\ Bernreuther \and E.\ A.\ M\"uller \and G.\ Jackson}
\authorrunning{Horsch \etal{}}
\institute{M.\ T.\ Horsch, S.\ K.\ Miroshnichenko, and J.\ Vrabec \at Thermodynamik und Energietechnik (ThEt), Institut f\"ur Verfahrenstechnik, Universit\"at Paderborn, Warburger Str.\ 100, 33098 Paderborn, Germany \and M.\ Bernreuther, C.\ W.\ Glass, and C.\ Niethammer \at H\"ochstleistungsrechenzentrum Stuttgart (HLRS), Nobelstr.\ 19, 70569 Stuttgart, Germany \and M.\ T.\ Horsch, G.\ Jackson, and E.\ A.\ M\"uller \at Molecular Systems Engineering (MSE), Centre for Process Systems Engineering, Imperial College London, South Kensington Campus, London SW7 2AZ, England}
%
%
\maketitle

\abstract{
Curved fluid interfaces are investigated on the nanometre length
scale by molecular dynamics simulation. Thereby, droplets
surrounded by a metastable vapour phase are stabilized in the canonical
ensemble. Analogous simulations
are conducted for cylindrical menisci separating vapour and liquid phases
under confinement in planar nanopores.
Regarding the emergence of nanodroplets during nucleation, a non-equilibrium
phenomenon, both the non-steady dynamics of condensation processes
and stationary quantities related to
supersaturated vapours are considered.
Results for the truncated and shifted Lennard-Jones fluid and for
mixtures of quadrupolar fluids confirm the applicability of the capillarity approximation
and the classical nucleation theory.
}

\section{Introduction}
\label{sec:intro}

The influence of curvature on the properties of a nanodroplet, as opposed to
an interface that is planar (on the molecular level), is hard
to capture experimentally. Yet it is important for refrigeration and energy technology
as well as meteorology to understand fluid interfaces with extremely high
curvatures because they characterize the onset of
condensation and boiling processes.

Beginning in the $1930$s with the work of \textsc{Verschaffelt} \cite{Verschaffelt36},
researchers became aware of the necessity of taking the internal structure of
fluid interfaces into account.
They increasingly looked beyond the picture of a
discrete dividing surface as postulated by Gibbs \cite{Gibbs78b}. In 
the subsequent years, this led to the theoretical work
of \textsc{Guggenheim} \cite{Guggenheim40} and \textsc{Tolman} \cite{Tolman49} which,
nonetheless, was mostly still based on effective radii and hence on
discretization.

\begin{figure}[t!]
\sidecaption
\centering
\includegraphics[width=\figwidth]{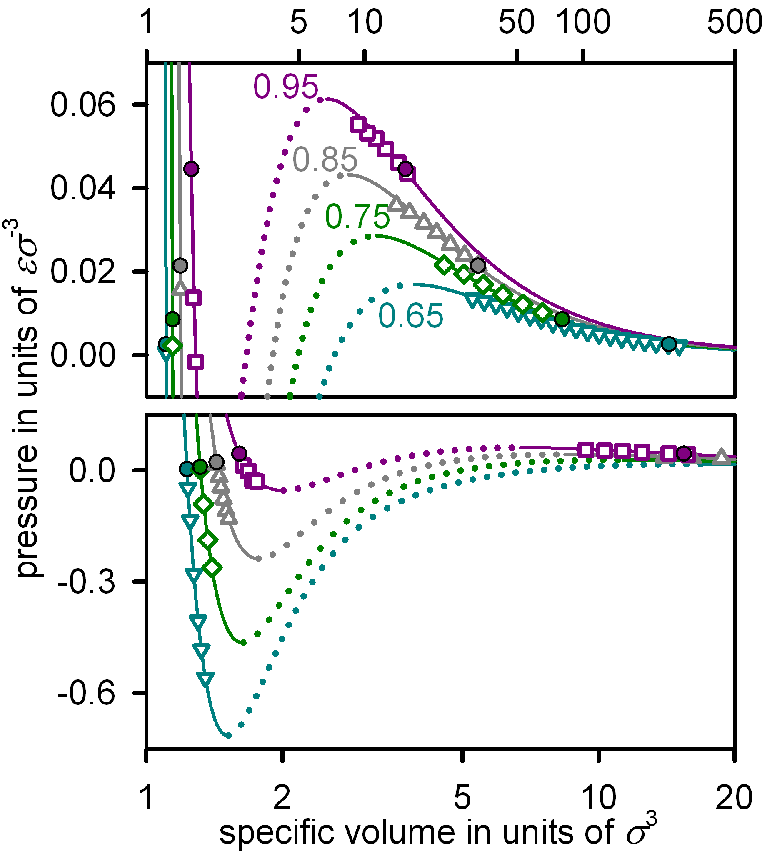}
\caption{
   \index{Lennard-Jones potential!Truncated-shifted}
   Isotherms in a pressure-volume diagram for the \LJTS{} fluid
   as determined by canonical MD simulation at temperatures
   of $\temperature$ = $0.65$ ($\triangledown$), $0.75$ ($\lozenge$),
   $0.85$ ($\triangle$), and $0.95$ $\LJenergy$ ($\square$) as well
   as the saturated states ($\bullet$) according to
   \textsc{Vrabec} \textit{et al.}\ \cite{VKFH06}
   in comparison with the present fifth-order virial expansion (---), cf.\ Tab.\
   \ref{tab:nukthevirial} and Eq.\ (\ref{eqn:leidenform}).
   Results in the vicinity of the spinodal line are
   not shown here, and no such values were used for adjusting the virial
   coefficients, to ensure that only states unperturbed
   by nucleation were taken into account.
}
\label{nukkin2}
\end{figure}

Today, molecular dynamics (MD) simulation provides a means of accessing the
internal structure and the non-equilibrium behaviour of
vapour-liquid interfaces directly, on the basis of physically sound but
algebraically simple effective pair potentials. For the truncated and shifted
Lennard-Jones (\LJTS) potential \cite{AT87}
\begin{equation}
   \potential(\distance) = \left\{
      \begin{array}{ll}
         4\LJenergy\left[\LJlength^{12}(\distance^{-12} - \distancecut^{-12})
            + \LJlength^{6}(\distancecut^{-6} - \distance^{-6})\right],
               & \textrm{\,\,for\,\,} \distance < \distancecut, \\
         0, & \textrm{\,\,for\,\,} \distance \geq \distancecut, 
      \end{array}
   \right.
\end{equation}
with a cutoff radius of $\distancecut$ = $2.5$ $\LJlength$, which constitutes a
reliable model for the noble gases and methane \cite{VKFH06},
fluid phase boundaries have
been simulated by several groups in the recent past \cite{VKFH06,
HL08, MPSF08, BDOVB10, HVH08}. Molecular simulation is particularly
suitable for investigating metastable states, cf.\ Fig.\ \ref{nukkin2}.
Virial isotherms that accurately describe the conditions
of fluid phase coexistence at planar or curved interfaces
\begin{equation}
   \frac{\pressure}{\temperature} = \sum_{\intb=1}^{5}\virialcoefficient_\intb\density^\intb,
\label{eqn:leidenform}
\end{equation}
which were correlated to the present data, are given in Tab.\ \ref{tab:nukthevirial}.
Therein, $\pressure$ is the pressure, $\temperature$ is the temperature,
$\density$ is the density, and the convention $k = 1$ is used.
On this basis,
the present work regards both equilibrium and non-equilibrium phenomena
for vapour-liquid interfaces of the \LJTS{} fluid. Nucleation
in supersaturated vapours is considered for mixtures of quadrupolar fluids
as well. For a more detailed exposition, the reader is pointed to four recent
articles \cite{HVH08, HV09b, HHDHHV10, HLWHV10}.

\section{Curved fluid interfaces in equilibrium: Theory}
\label{sec:thcur}

The Tolman \cite{Tolman49} approach to curved vapour-liquid interfaces
is based on comparing different effective radii of a droplet,
namely the Gibbs adsorption radius $\radequimolar$, for which the interfacial
excess density is zero, and the Laplace radius $\radlaplace$ from
\begin{equation}
   \surfacetension = \frac{1}{2}\radlaplace(\pressure_\inta - \pressure),
\label{eqn:pLaplace}
\end{equation}
i.e.\ the Laplace equation in terms of the surface tension $\surfacetension$
and the pressure $\pressure_\inta$ inside a droplet containing $\inta$
molecules under equilibrium conditions. The deviation between these radii,
the Tolman length
\begin{equation}
   \tolmanlength = \radequimolar - \radlaplace,
\end{equation}
characterizes the curvature dependence of the surface
tension \cite{Tolman49}. The \textit{capil\-larity approximation}, which postulates
the droplets to be spherical and incompressible with a curvature independent
surface tension, assumes $\tolmanlength$ to be zero.

\begin{table}[h!]
\centering
\caption{
   \label{tab:nukthevirial}
   Virial coefficients for the \LJTS{} fluid as determined from a fit to the MD simulation
   results shown in Fig.\ \ref{nukkin2} and to the saturated vapour and liquid densities
   determined by Vrabec \etal{} \cite{VKFH06}. The spinodal densities $\spinodal{(\liq\density)}$
   and $\spinodal{(\vap\density)}$ for liquid and vapour, respectively, were determined from
   the virial expansion, cf.\ Eq.\ (\ref{eqn:leidenform}).
}
\begin{tabular}{c||cccc|cccc}
   $\temperature$ & $- \virialcoefficient_2$ & $\virialcoefficient_3$ & $- \virialcoefficient_4$ & $\virialcoefficient_5$
                  & $\liq\density$ & $\vap\density$ & $\spinodal{(\liq\density)}$ & $\spinodal{(\vap\density)}$
   \\ \hline
   $0.65$ & $11.7675\phantom{0}$ & $44.5866\phantom{0}$ & $96.9625$ & $71.4351$
          & $0.813$ & $0.00406$ & $0.660$ & $0.0592$ \\
   $0.7\phantom{0}$ & $\phantom{0}9.77572$ & $34.176\phantom{00}$ & $76.4866$ & $59.4954$
                    & $0.787$ & $0.00728$ & $0.636$ & $0.0740$ \\
   $0.75$ & $\phantom{0}8.43697$ & $27.7315\phantom{0}$ & $62.373\phantom{0}$ & $50.3464$
                    & $0.759$ & $0.0124\phantom{0}$ & $0.613$ & $0.0886$ \\
   $0.8\phantom{0}$ & $\phantom{0}7.33394$ & $21.854\phantom{00}$ & $41.1349$ & $40.3329$
                    & $0.730$ & $0.0198\phantom{0}$ & $0.588$ & $0.103\phantom{0}$ \\
   $0.85$ & $\phantom{0}6.48592$ & $18.3318\phantom{0}$ & $40.0252$ & $34.6962$
                    & $0.699$ & $0.0304\phantom{0}$ & $0.564$ & $0.119\phantom{0}$ \\
   $0.9\phantom{0}$ & $\phantom{0}5.44587$ & $12.3036\phantom{0}$ & $25.0989$ & $23.6305$
                    & $0.664$ & $0.0446\phantom{0}$ & $0.532$ & $0.134\phantom{0}$ \\
   $0.95$ & $\phantom{0}4.97043$ & $10.0411\phantom{0}$ & $17.1387$ & $16.0653$
                    & $0.622$ & $0.0648\phantom{0}$ & $0.499$ & $0.149\phantom{0}$ \\
   $1\phantom{.00}$ & $\phantom{0}4.67665$ & $\phantom{0}9.83155$ & $15.6063$ & $13.8778$
                    & $0.571$ & $0.0962\phantom{0}$ & $0.466$ & $0.174\phantom{0}$
\end{tabular}
\end{table}

According to the formalism employed by Buff \cite{Buff55}
and Kondo \cite{Kondo56}, the surface tension becomes mimimal if it
is evaluated with respect to $\radlaplace$. It can be shown that
this assertion is only valid if the interfacial area $\area$ is
proportional to $\radlaplace^2$. However, both mechanical and
thermodynamic equilibrium conditions for a droplet containing $\inta$ molecules imply
\begin{equation}
   \radlaplace = 2 \left(\frac{\partial\volume_\inta}{\partial\area}\right)_{\absnum, \volume, \temperature},
\end{equation}
where $\volume_\inta$ and $\volume$ are the volumes occupied by the
droplet and the system as a whole, respectively.
This only agrees with $\area \sim \radlaplace^2$ if curvature effects
cancel out.

For cylindrical interfaces, the surface tension varies with the radius according to
\begin{equation}
   \left[\left(\frac{\partial\ln\radlaplace}{\partial\ln\surfacetension}\right)_\temperature - 1\right]^{-1}
      = \frac{\tolmanlength}{\radlaplace} + \frac{\tolmanlength^2}{2\radlaplace^2},
\label{eqn:tolmancyl}
\end{equation}
an expression that is similar and analogous to Tolman's equation for droplets, where
$\radlaplace$ is defined to be positive for convex and negative for concave menisci.
By combining the Young equation \cite{Young05}
with an expansion of $\surfacetension$ to first order in $1\slash\radlaplace$,
the contact angle $\contactangle$ of a fluid confined in a planar nanopore is obtained as
\begin{equation}
   \cos\contactangle = \left(\frac{\planartension}{\Dgs} + \frac{\tolmaninf}{\radmin}\right)^{-1}.
\label{eqn:youngtolman}
\end{equation}
Therein, $\planartension$ is the surface tension of the planar vapour-liquid interface,
$\tolmaninf$ is the Tolman length in the planar limit,
and the pore diameter is $2\radmin$, while $\Dgs$ indicates the difference between the
specific surface energies of the substrate when it is in contact with the vapour and
the liquid, respectively.

\section{Curved fluid interfaces in equilibrium: MD simulation}
\label{sec:mdcur}

\begin{figure}[b!]
\sidecaption
\centering
\includegraphics[width=\figwidth]{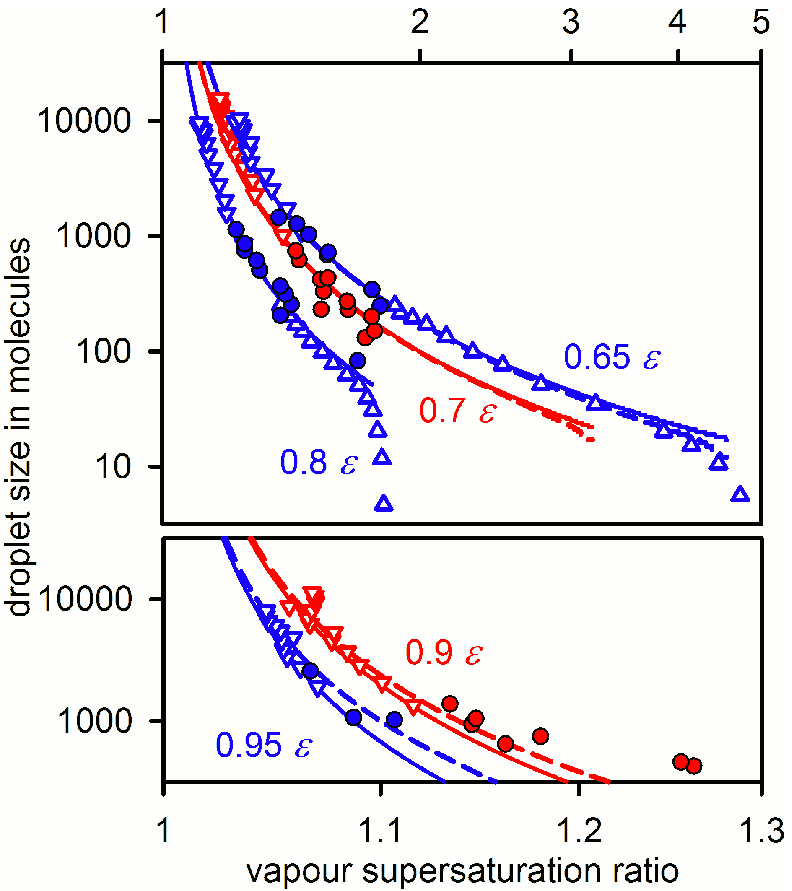}
\caption{
   Droplet size $\critical\nuclsize$ over
   the supersaturation ratio $\supersatmu$ (in terms of the chemical potential)
   for vapour-droplet equilibria of the \LJTS{} fluid
   ($\bullet$) from the present work,
   ($\triangledown$) according to Vrabec \etal{} \cite{VKFH06},
   ($\triangle$) according to \textsc{Napari} \textit{et al.}\ \cite{NJV09},
   and following the capillarity approximation (---)
   as well as the Laaksonen \etal{} \cite{LFK94} model ($\textnormal{-- --}$).
}
\label{nukdro1}
\end{figure}

Using the \mardyn{} MD program \cite{BV05}, equilibrium states involving
droplets and cylindrical menisci were simulated for the \LJTS{} fluid.

\begin{figure}[h!]
\sidecaption
\centering
\includegraphics[width=\figwidth]{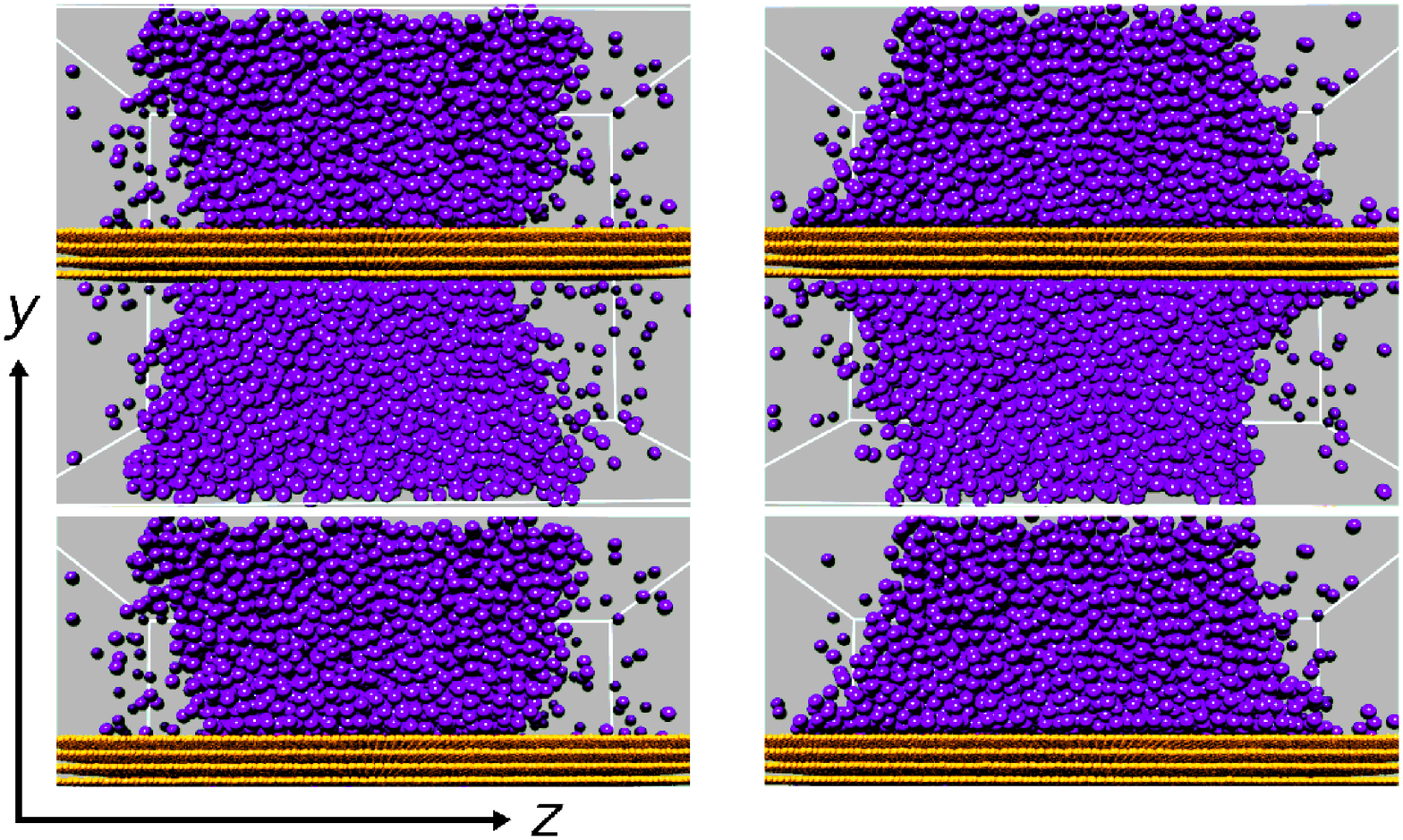}
\caption{Simulation snapshots for the reduced fluid-wall dispersive energy $\FWenergy$
of $0.09$ (left) and $0.16$ (right) at a temperature of $0.73$ $\LJenergy$.
The upper half is reproduced in the bottom to illustrate the effect of the periodic
boundary condition.}
\label{fwcmen1}
\end{figure}

Vapour-droplet equilibrium MD simulations
were conducted for droplets containing on the order of $100$ to $1$ $000$ molecules
in the canonical ensemble, where such equilibria can be stable -- as opposed e.g.\
to the grand canonical ensemble where this corresponds to a free energy
maximum. The droplet size was evaluated according to a version of the cluster criterion of
\textsc{ten Wolde} and \textsc{Frenkel} \cite{TF98}
with the connectivity radius $\radstillinger = 1.5$
$\LJlength$ and coordination numbers $\intb \geq 4$ defining the liquid phase. The present results
correspond to moderately supersaturated vapours, cf.\ Fig.\ \ref{nukdro1},
and are consistent with the results of Vrabec \etal{} \cite{VKFH06} on larger droplets
as well as the study of Napari \etal{} \cite{NJV09} covering
vapours at pressures that approach the spinodal line.
In the intermediate regime, the droplet size in equilibrium
generally agrees well with the capillarity approximation.

\begin{figure}[t!]
\sidecaption
\centering
\includegraphics[width=\figwidth]{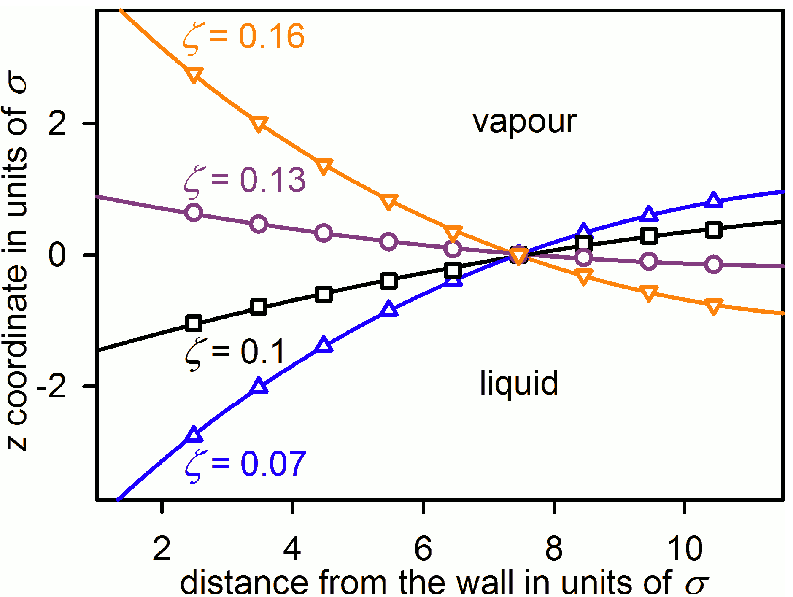}
\caption{Vapour-liquid interface profiles for the reduced fluid-wall dispersive
energy $\FWenergy$ of $0.07$ ($\triangle$), $0.10$ ($\square$), $0.13$ ($\circ$), and
$0.16$ ($\triangledown$) at a temperature of 0.82 $\LJenergy$.}
\label{fwcmen2}
\end{figure}

\begin{figure}[h!]
\sidecaption
\centering
\includegraphics[width=\figwidth]{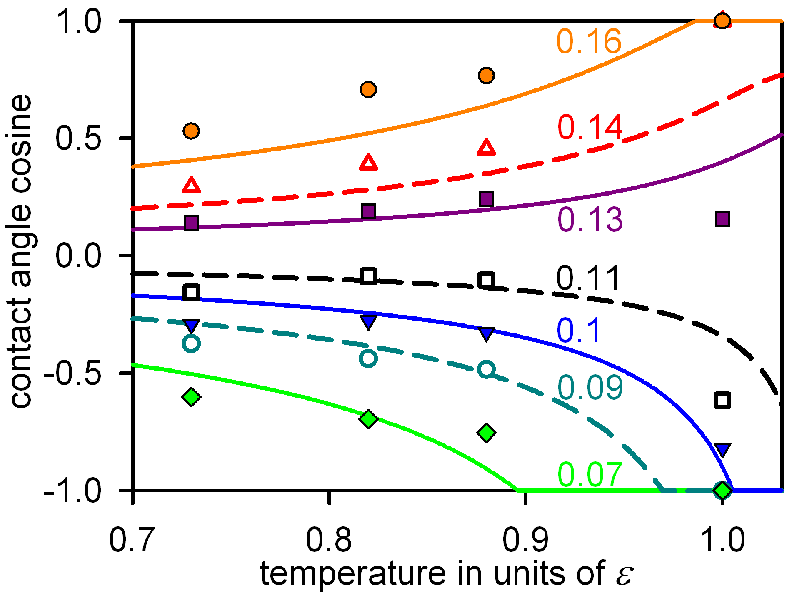}
\caption{MD simulation results for $\contactangle$ over $\temperature$
with a reduced fluid-wall dispersive energy of $\FWenergy$ = $0.07$ ($\blacklozenge$),
$0.09$ ($\circ$), $0.1$ ($\blacktriangledown$), $0.11$ ($\square$),
$0.13$ ($\blacksquare$), $0.14$ ($\triangle$), and $0.16$ ($\bullet$)
as well as following the proportionality law (---,\, $\textnormal{-- --}$), cf.\
Eq.\ (\ref{eqn:youngtolman}), with $\FWrect$ = $0.118$, $\tolmaninf(\temperature)$ from
a correlation based on the data of Vrabec \etal{} \cite{VKFH06}, and a reduced
fluid-wall surface energy difference of $\redDgs$ = $7$
$\LJlength\LJenergy$, cf.\ Eq.\ (\ref{eqn:redDgs}).}
\label{fwcmen4}
\end{figure}

Cylindrical interfaces were investigated by simulating liquid slabs, cf.\ Fig.\ \ref{fwcmen2},
confined between two planar and layered walls represented by coupled harmonic
oscillators. The equilibrium positions of the wall atoms were
aligned according to a hexagonal structure with an interatomic distance of $0.3816$ $\LJlength$,
corresponding to the bond length in graphite expressed in terms of the $\LJlength$
parameter value for fluid methane.
Both the fluid-fluid and the fluid-wall interactions were modelled by
the \LJTS{} potential, employing equal size parameters $\LJlengthFW = \LJlength$
in both cases, while the dispersive energy between fluid
molecules and wall atoms
\begin{equation}
   \LJenergyFW = \FWenergy\LJenergy,
\end{equation}
was systematically varied. The arithmetic mean of the
saturated vapour and liquid densities
was selected as a criterion for detecting the phase boundary.
A circle was adjusted to the resulting profile
at distances between $2$ and $11$ $\LJlength$ from the wall,
cf.\ Fig.\ \ref{fwcmen2},
and the tangent to this circle at a distance of $1$ $\LJlength$ from
the wall was examined to determine the contact angle.
Quali\-tatively, the contact angles obtained by the present MD
simulations are captured by Eq.\ (\ref{eqn:youngtolman}), assuming a
proportionality law for
\begin{equation}
   \Dgs = \redDgs (\liq\density - \vap\density) (\FWenergy - \FWrect),
\label{eqn:redDgs}
\end{equation}
cf.\ Fig.\ \ref{fwcmen4}.
The magnitude of the fluid-wall dispersion
for which $\contactangle$ becomes rect\-angular ($\FWrect = 0.118$)
was found to be temperature independent.

\section{Homogeneous vapour to liquid nucleation: Theory}
\label{sec:thnuc}

\begin{figure}[b!]
\sidecaption
\centering
\includegraphics[width=\figwidth]{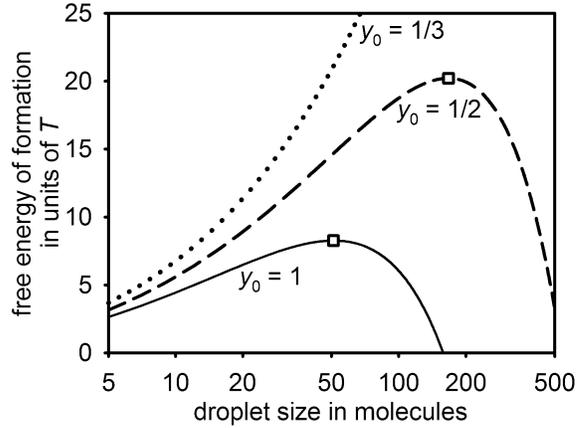}
\caption{
   Free energy of formation according to CNT for
   \chem{CO_2} droplets in supersaturated vapours
   at $\temperature$ = $250.2$ K with $\supersatrho$ = $2.72$
   and \chem{CO_2} mole fractions of $\supfraction$ = $1$ (---),
   $1/2$ ($\textnormal{-- --}$), and $1/3$ ($\cdots$), reaching a maximum
   for the critical droplet ($\square$). Note that in the thermodynamic limit, i.e.\
   in a macroscopic system, the \textsc{Gibbs}, \textsc{Helmholtz}, and \textsc{Landau} free energies
   of formation for small droplets converge.
}
\label{phdCO2air1}
\end{figure}

The foundations of the classical nucleation theory (CNT),
concerning the first step of a first-order phase transition
in the bulk of a metastable phase, were laid by Volmer
and Weber \cite{VW26} as well as Farkas \cite{Farkas27}.
On the basis of the capillarity approximation,
the free energy of formation $\Delta\helmholtz$ of a droplet containing $\inta$
molecules in the thermodynamic limit (i.e.\ for an infinitely large vapour phase
at a constant supersaturation ratio) evaluates to
\begin{equation}
   \frac{\partial\helmholtz}{\partial\inta}
      = \planartension\frac{\partial\area}{\partial\inta}
         - \left(\chempot - \chempot_\inta
            + \frac{\pressure_\inta - \pressure}{\liq\density}\right),
\label{eqn:dropletA}
\end{equation}
in differential terms, where $\chempot_\inta$ and $\pressure_\inta$ are the
chemical potential and the pressure inside the droplet, respectively, while
$\chempot$ and $\pressure$ refer to the vapour. 
In the $\inta\to\infty$ limit, the rightmost term of Eq.\ (\ref{eqn:dropletA})
yields an effective chemical potential difference
\begin{equation}
   \Deltamueff = \chempot - \sat\chempot(\temperature) + \frac{\sat\pressure(\temperature) - \pressure}{\liq\density},
\label{eqn:Deltamueff}
\end{equation}
that accounts for the \qq{pressure effect} \cite{WHBR08} of the vapour -- which may include
the contribution of an inert carrier gas. Note that
$\sat\chempot(\temperature)$ and $\sat\pressure(\temperature)$
do not depend on $\inta$ since these quantities characterize
the saturated bulk fluid. As visualized in Fig.\ \ref{phdCO2air1},
the presence of a carrier gas increases the free
energy barrier $\critical{\Delta\helmholtz}$ of a nucleation process, i.e.\ the
maximum of $\Delta\helmholtz$ reached for a \textit{critical droplet} in (unstable) equilibrium
with the vapour, corresponding to the conditions discussed above.

\begin{figure}[h!]
\sidecaption
\centering
\includegraphics[width=\figwidth]{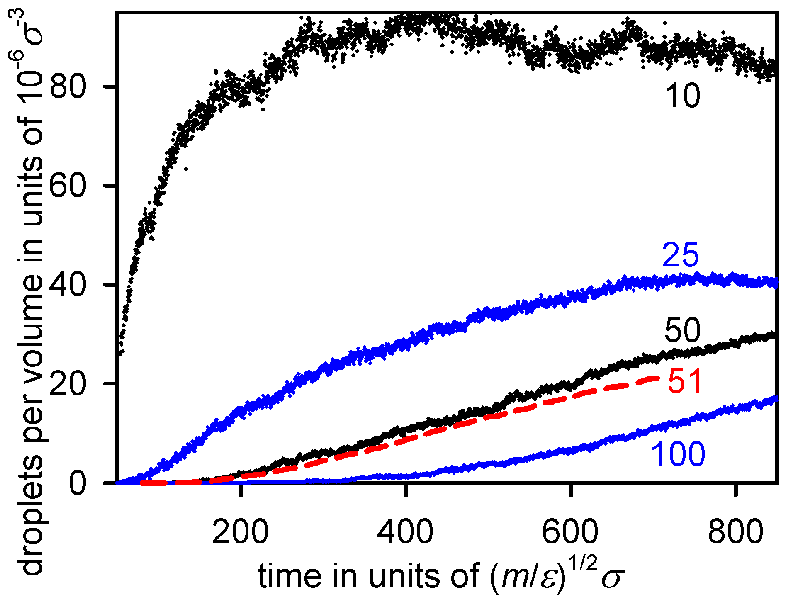}
\caption{
   Number of droplets per volume over simulation time for droplets
   containing $\inta > 10$, $25$, $50$, and $100$ molecules in a canonical
   ensemble MD simulation of the \LJTS{} fluid at $\temperature$ =
   $0.7$ $\LJenergy$ and $\density$ = $0.03421$ $\LJlength^{-3}$ 
   in comparison with the aggregated number of \demon{}
   interventions per volume in a grand canonical MD simulation with
   $\temperature$ = $0.7$ $\LJenergy$, $\supersatmu$ = $2.8658$,
   and $\threshold$ = $51$.
}
\label{nukuVT1}
\end{figure}

From the analysis of a random walk over $\inta$, the probability for
a droplet containing $\threshold$ molecules to eventually reach
macroscopic size can be determined as
\begin{equation}
   \probinftyof{\threshold}
      = \frac{\int_1^\threshold
         \exp\left(2 \Delta\helmholtz\slash\temperature\right) \differential\nuclsize
            }{ \int_1^\infty
	       \exp\left(2 \Delta\helmholtz\slash\temperature\right) \differential\nuclsize
	       },
   \label{eqn:intervention}
\end{equation}
while the nucleation rate, i.e.\ the number of macroscopic liquid drops formed
by homogeneous nucleation per volume and time unit, is
\begin{equation}
   \nuclrate = \CNTpreexp\nonaccommodation \collrate \cdot
      \frac{\monomers\zeldovich\critical{\area}}{\volume}
         \exp\left(\frac{-\critical{\Delta\helmholtz}}{\temperature
            }\right),
\label{eqnFRLP66}
\end{equation}
according to CNT. In this expression, $\collrate$ refers to the rate at which
vapour monomers collide with an interface (per surface area), $\monomers$ is the
number of monomers in the system, $\critical{\area}$ is the surface area of
a critical droplet, and $\volume$ is the system volume. The correction
factors $\CNTpreexp$, $\nonaccommodation$, and $\zeldovich$ were introduced by
Farkas \cite{Farkas27}, Feder \etal{} \cite{FRLP66}, and
Zel'dovi\v{c} \cite{Zeldovich42}, respectively. The overall pressure effect
on $\nuclrate$ following Eq.\ (\ref{eqnFRLP66}), as
discussed by Wedekind \etal{} \cite{WHBR08}, defines the carrier gas
correction factor $\Weffect$.

\section{Homogeneous vapour to liquid nucleation: MD simulation}
\label{sec:mdnuc}

Nucleation in supersaturated vapours was studied by simulating
systems containing between $100$ $000$ and $17$ $000$ $000$ molecules,
exploiting the excellent scalability of the \mardyn{} program on massively
parallel computing platforms \cite{BBV10}.
%
%

\begin{figure}[t!]
\sidecaption
\centering
\includegraphics[width=\figwidth]{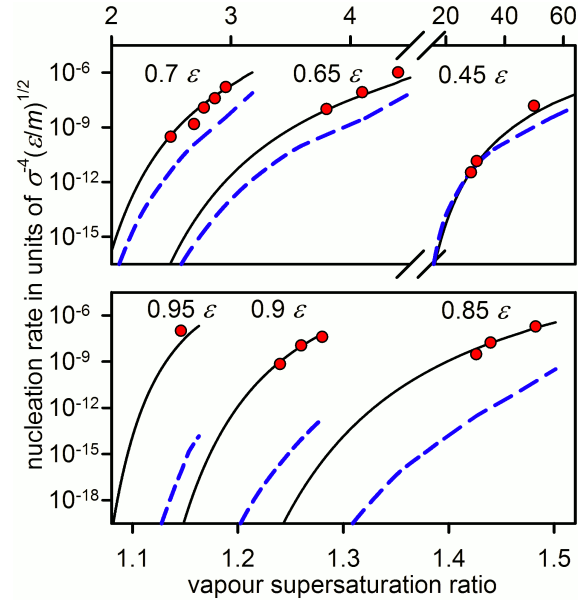}
\caption{
   Nucleation rate of the \LJTS{}
   fluid over supersaturation from the present MD
   simulations of the grand canonical ensemble
   with \textsc{McDonald}'s \demon{} ($\bullet$)
   as well as according to CNT with $\CNTpreexp
   = 200$ (---) and
   the Laaksonen \etal{} \cite{LFK94} model ($\textnormal{-- --}$)
   at temperatures of $\temperature$ = $0.45$, $0.65$, $0.7$, $0.85$, $0.9$, and $0.95$ $\LJenergy$.
   The supersaturation ratio is given in terms of the chemical potential
   of the vapour phase.
}
\label{nukuVT3}
\end{figure}

The method of Yasuoka and
Matsumoto \cite{YM98}, where droplet formation rates are evaluated during the
stage of a condensation process that corresponds to nucleation (rather than
relaxation or droplet growth), was applied to
the canonical ensemble. In these simulations, the vapour pressure decreased over time
due to the transfer of molecules from the vapour to the dispersed liquid phase.
Furthermore, steady state quantities, pertaining to nucleation only, were
investigated with a new simulation method. This method combines the grand
canonical ensemble with \textit{McDonald's d\ae{}mon} \cite{McDonald63},
an intelligent being whose interventions eliminate droplets containing more
than $\threshold$ molecules; see Fig.\ \ref{nukuVT1} for a comparison between
these approaches. Results for the \LJTS{} fluid agree well with CNT,
using a temperature independent value for the empirical correction factor $\CNTpreexp$
introduced by Farkas \cite{Farkas27}.

\begin{table}[h!]  
\centering
\caption{
   \label{tab:nukpre2}
   Droplet formation rate from Yasuoka-Matsumoto (YM) canonical
   ensemble MD simulation as well as critical droplet size (in
   molecules), \textsc{Wedekind} factor $\Weffect$, and the prediction 
   $\CNT{\nuclrate\slash\committor}(\threshold)$
   for the droplet formation rate according to CNT with $\CNTpreexp$ = $1$,
   in dependence of temperature (in units of K), supersaturation ratio (with
   respect to the partial density of carbon dioxide) and YM threshold size $\threshold$ (in
   molecules) for the quaternary system $\chem{CO_2} + \chem{N_2} + \chem{O_2}
   + \chem{Ar}$. The mole fraction $\supfraction$ of carbon dioxide in the supersaturated
   vapour is indicated in the table, while the composition regarding
   the other fluids corresponds to the earth's atmosphere. The rates are given in units of m$^{-3}$s$^{-1}$
   and where no nucleation was detected, $\CNT{\nuclrate}$ instead of $\CNT{\nuclrate\slash\committor}$
   is shown in the last column.
}
\begin{tabular}{cccc||c|clc}
$\temperature$ & $\density\slash\vapor\density$ & $\supfraction$
                           & $\threshold$
                           & $\nuclrateof{\threshold}$
                           & $\NcritCNT$ & $\Weffect$ 
                           & $\CNT{\nuclrate\slash\committor}$ \\ \hline
$238.4$ & $2.80$ & $1/2$ & $\phantom{0}50$
      & $1.5 \mult 10^{33}$ 
      & $\phantom{0}66$ & $0.03$ 
      & $2.6 \mult 10^{31}$ \\
      &      & & $\phantom{0}85$
      & $1.6 \mult 10^{32}$ 
      & & 
      & $3.3 \mult 10^{30}$ \\
      &      & $1\phantom{/1}$ & $\phantom{0}50$

      & $5.6 \mult 10^{32}$ 
      & $\phantom{0}41$ &  $1$
      & $9.9 \mult 10^{31}$ \\
      &      &   & $\phantom{0}85$        
      & $2.1 \mult 10^{32}$ 
      & & 
      & $7.6 \mult 10^{31}$ \\
      & $3.08$ & $1/2$ & $\phantom{0}50$
      & $5.5 \mult 10^{33}$ 
      & $\phantom{0}65$ & $0.02$
      & $3.1 \mult 10^{31}$ \\
      &      & & $150$         
      & $3.1 \mult 10^{32}$ 
      & & 
      & $3.9 \mult 10^{30}$ \\
      &      & $1\phantom{/1}$ & $\phantom{0}50$
      & $6.3 \mult 10^{33}$ 
      & $\phantom{0}39$ & $1$
      & $1.6 \mult 10^{32}$ \\
      &      & & $150$          
      & $2.9 \mult 10^{32}$ 
      & & 
      & $1.3 \mult 10^{32}$ \\
      & $3.36$ & $1/3$ & ---
      & $\ll 10^{31}$ 
      & $127$ & $4.2\mult10^{-6}$
      & $1.1 \mult 10^{27}$ \\
      &      & $1/2$ & $\phantom{0}50$
      & $1.1 \mult 10^{34}$ 
      & $\phantom{0}65$ & $0.02$
      & $8.7 \mult 10^{30}$ \\
      &      & & $300$          
      & $3.2 \mult 10^{32}$ 
      & & 
      & $4.2 \mult 10^{30}$ \\
      &      & $1\phantom{/1}$ & $\phantom{0}50$
      & $6.7 \mult 10^{33}$ 
      & $\phantom{0}37$ & $1$
      & $2.1 \mult 10^{32}$ \\
      &      & & $300$         
      & $1.4 \mult 10^{33}$ 
      & & 
      & $1.8 \mult 10^{32}$ \\ \hline
$250.2$ & $2.34$ & $1/2$ & $\phantom{0}50$
      & $1.1 \mult 10^{34}$ 
      & $140$ & $1.9\mult10^{-4}$
      & $1.8 \mult 10^{33}$ \\
      &      & & $100$          
      & $1.1 \mult 10^{33}$ 
      & & 
      & $7.8 \mult 10^{29}$ \\
      &      & $1\phantom{/1}$ & $\phantom{0}50$
      & $1.3 \mult 10^{33}$ 
      & $\phantom{0}54$ & $1$
      & $3.9 \mult 10^{32}$ \\
      &      & & $100$          
      & $3.4 \mult 10^{32}$ 
      & & 
      & $1.4 \mult 10^{32}$ \\
      & $2.53$ & $1/2$ & $\phantom{0}85$
      & $7.4 \mult 10^{33}$ 
      & $143$ & $1.0 \mult 10^{-4}$
      & $3.9 \mult 10^{30}$ \\
      &      & & $200$          
      & $7.4 \mult 10^{32}$ 
      & & 
      & $3.1 \mult 10^{28}$ \\
      &      & $1\phantom{/1}$ & $\phantom{0}85$
      & $2.2 \mult 10^{33}$ 
      & $\phantom{0}52$ & $1$
      & $1.9 \mult 10^{32}$ \\
      &      & & $200$          
      & $7.7 \mult 10^{32}$ 
      & & 
      & $1.9 \mult 10^{32}$ \\
      & $2.72$ & $1/3$ & ---
      & $\ll 10^{31}$ 
      & $879$ & $4.3 \mult 10^{-25}$
      & $2.3 \mult 10^{8\phantom{0}}$ \\
      &        & $1/2$ & $\phantom{0}75$
      & $1.3 \mult 10^{34}$ 
      & $150$ & $4.2\mult10^{-5}$
      & $1.8 \mult 10^{31}$ \\
      &      & & $250$          
      & $1.6 \mult 10^{33}$ 
      & & 
      & $1.7 \mult 10^{28}$ \\
      &      & $1\phantom{/1}$ & $\phantom{0}75$
      & $4.8 \mult 10^{33}$ 
      & $\phantom{0}50$ & $1$
      & $2.6 \mult 10^{32}$ \\
      &      & & $250$          
      & $1.4 \mult 10^{33}$ 
      & & 
      & $2.5 \mult 10^{32}$ \\ \hline
\end{tabular}
\end{table}

%
Canonical ensemble MD simulations were also conducted for multi-component
systems containing nitrogen, oxygen, and argon -- at the ratio prevalent in the earth's
atmosphere -- as well as carbon dioxide with a greater partial density than at saturation.
The molecular models employed for this purpose, introduced by Vrabec \etal{} \cite{VSH01},
are well-established with respect to fluid phase equilibria \cite{VSH01, HVH09,
VKBMH09}. For these systems, the analysis of the carrier gas effect according to Wedekind \etal{} \cite{WHBR08}
is confirmed quali\-tatively by the determined droplet formation rates \nuclrateof{\threshold},
given in Tab.\ \ref{tab:nukpre2}, although significant quantitative deviations
are present at high temperatures.

\section{Conclusion}
\label{sec:concl}

From the preceding analysis of curved vapour-liquid interfaces and homogeneous
nucleation it can be concluded that CNT is able to capture both the nucleation
rate and the critical droplet size for the considered systems, i.e.\ the \LJTS{}
fluid and a quaternary mixture of quadrupolar and unpolar fluids.
The main criticism usually made of CNT is that it applies the
capillarity approximation to small droplets where significant curvature
effects should be expected. However, a deviation from capillarity
is implicit in the prefactor $\CNTpreexp$ which empirically
accounts for its overall influence on the nucleation rate. This corresponds to stating
that the capillarity approximation overestimates the free energy
barrier by $\temperature\ln\CNTpreexp$.

The physical foundation of this approach is more robust than it might
seem at first sight. By combining recent simulation results on the equilibrium
vapour pressure of droplets, cf.\ Fig.\ \ref{nukdro1}, it becomes apparent that curvature
effects are signi\-ficant in the immediate vicinity of the spinodal line for the vapour,
corresponding to $\critical\nuclsize < 100$, while they are
virtually undetectable for droplets containing more than $1$ $000$ molecules.
Thus, the deviation from Eq.\ (\ref{eqn:dropletA}) regarding the magnitude of
$\critical{\Delta\helmholtz}$ is dominated by an integral over
the free energy of formation for \textit{extremely small} droplets. At supersaturation ratios
sufficiently distant from spinodal conditions, this contribution does not depend
on $\supersat$ and can be represented by $-\temperature\ln\CNTpreexp$ with
a constant value of $\CNTpreexp$.

\smallskip

\textbf{Acknowledgement.}
The authors would like to thank J.\
Harting, H.\ Hasse, E.\ Y.\ Kenig, and G.\ Reina
for their support and for valuable discussions.
The present work, which contributes to the BMBF project IMEMO, was
conducted under the auspices of the Boltzmann-Zuse Society of
Computational Molecular Engineering (BZS). Major aspects of it
were facilitated by the reliable technical assistance
of M.\ Heitzig and Z.\ Lin.
The position of M.\ T.\ Horsch at the Imperial College London is
funded by the DAAD postdoc programme,
and computations were performed at the High Performance
Computing Center Stutt\-gart (HLRS) with resources assigned
to the grant MMHBF.

\end{document}